\documentclass{IEEEtran}
\usepackage{cite}
\usepackage{amsmath,amssymb,amsfonts}
\usepackage{mathtools}
\usepackage{algorithmic}
\usepackage{graphicx}
\usepackage{stfloats}
\usepackage{textcomp}
\usepackage{tablefootnote}
\usepackage[caption=false,font=footnotesize]{subfig}
\usepackage{placeins}
\newcommand{\diagop}{\mathrm{diag}}
\newcommand{\vecop}{\mathrm{vec}}
\usepackage{colortbl}
\usepackage{xcolor}
\usepackage{float}
\def\BibTeX{{\rm B\kern-.05em{\sc i\kern-.025em b}\kern-.08em
    T\kern-.1667em\lower.7ex\hbox{E}\kern-.125emX}}
\begin{document}
\title{Joint Target-Tag Estimation in Ambient Backscatter-Based ISAC}
\author{Awadhesh Gupta,~\IEEEmembership{Graduate Student Member,~IEEE}, 
Kumar Vijay Mishra,~\IEEEmembership{Senior Member,~IEEE}, \\
Naveen K. D. Venkategowda,~\IEEEmembership{Member,~IEEE}, 
and Aditya K. Jagannatham,~\IEEEmembership{Senior Member,~IEEE}
\thanks{The work of Aditya K. Jagannatham was supported in part by the Qualcomm Innovation Fellowship; in part by the Qualcomm 6G UR Gift; in part by the Arun Kumar Chair Professorship; in part by Anusandhan National Research Foundation's Advanced Research Grant ANRF/ARG/2025/005895/ENS; and in part by the funding support provided to ICON-project by DST and UKRI-EPSRC under India-UK Joint opportunity in Telecommunications Research.

A.G. and A. K. J. are with the Department of Electrical Engineering, Indian Institute of Technology Kanpur, 208016 India, e-mail: \{awadheshg20, adityaj\}@iitk.ac.in.}
\thanks{K. V. M. is with the United States DEVCOM Army Research Laboratory, Adelphi, MD 20783 USA, e-mail: kvm@ieee.org.}
\thanks{N. K. D. V. is with the Department of Science and Technology, Link\"{o}ping University, Norrk\"{o}ping, 60174 Sweden, email: naveen.venkategowda@liu.se.

}
}

\maketitle

\begin{abstract}
Integrated sensing and communication (ISAC) has emerged as a key paradigm for unifying radar sensing and wireless communications within a shared hardware and spectral framework. In parallel, ambient backscatter communication (AmBC) has gained attention as a low-power solution for enabling passive connectivity by modulating and reflecting existing radio-frequency signals. Motivated by these developments, this paper investigates an integrated sensing and ambient backscatter communication (ISABC) framework in which a dual-functional radar–communication (DFRC) transceiver jointly performs high-resolution target sensing and low-power tag detection in cluttered environments. Within this ISAC and AmBC setting, we develop a comprehensive system model that jointly captures the estimation of target and clutter parameters, including delay, Doppler shift, and complex reflectivity, together with the recovery of embedded AmBC tag symbols. The resulting DFRC signal model reveals strong coupling among direct-path reflections, clutter echoes, and passive backscatter signals, leading to a highly entangled estimation problem. To address this challenge, we propose a robust three-stage signal processing framework that sequentially estimates target and clutter parameters before detecting the AmBC tag message. This structured approach effectively disentangles the superimposed components, improving both sensing accuracy and communication reliability in ISAC systems. Extensive simulations, benchmarked against the Cramér–Rao lower bound (CRLB), demonstrate the superior estimation performance of the proposed DFRC-based solution. The results further highlight the fundamental trade-offs between sensing accuracy and backscatter communication efficiency in ISABC systems.
\end{abstract}


\section{Introduction}
Ambient backscatter communication (AmBC) and integrated sensing and communication (ISAC) are emerging as key enablers of future low-power and intelligent wireless systems \cite{ISAC_AmBC_IoT_Mag,arxiv_ISABC_article, ISAC_JSTEAP, Inverse_scattering_JSTEAP, Inband_ISAC_JSTEAP, ISABC_FAS}. ISAC enables simultaneous communication and sensing over shared time-frequency resources, significantly enhancing spectral efficiency, reducing hardware costs, and minimizing electromagnetic pollution \cite{ISAC_Vijay, Vijay_mmWave_ISAC, Vijay_mmwave_JRC_SP, Precipitation_sensing_vijay_JSTEAP, JRC_VJ_Sir_suggested}. This facilitates novel services such as high-precision localization, activity recognition, gesture tracking, and environmental mapping, while improving communication through more accurate beamforming and reduced channel estimation overhead \cite{MTT_VJSir_suggested, Onebit_iot_localization,3D_BL,data_aided, BSA_ISAC, OAMVortex_VJSIR_suggested}. Moreover, the joint utilization of sensing and communication functionalities enables efficient resource sharing, which is essential for next-generation wireless networks.

Backscatter communication (BC) systems typically consist of a carrier emitter (CE), a reader, and a BD/tag, and can be broadly classified into monostatic, bistatic, and ambient configurations. In monostatic BC systems, the CE and reader are co-located, often sharing the same antenna infrastructure, which leads to round-trip path loss and requires full-duplex operation. In contrast, bistatic BC systems spatially separate the CE and reader, offering improved deployment flexibility and reduced path loss. AmBC further extends this paradigm by eliminating the need for a dedicated CE and instead leveraging existing ambient RF sources, such as cellular base stations, Wi-Fi access points, or television transmitters. In AmBC systems, a backscatter device (BD), also referred to as a tag, harvests energy from ambient radio-frequency (RF) signals and conveys information by modulating and reflecting the incident waveform toward an access point (AP) or reader without generating its own RF signal \cite{Van_AmBC_Survey,Griffin_disadvantage_backscatter_2}. Information transmission is realized through impedance switching, where the tag alternates its reflection coefficient between discrete states (e.g., $+1$ and $-1$), corresponding to different information bits by varying the energy harvesting and reflection coefficients of its operating circuit. The AP subsequently decodes the transmitted information by exploiting the distinct signal characteristics associated with the tag's reflection states. Owing to their battery-free operation, low hardware complexity, and minimal maintenance requirements, AmBC systems have emerged as an attractive solution for energy-efficient IoT applications.

AmBC complements ISAC by enabling ultra-low-power, batteryless devices to communicate by modulating and reflecting ambient RF signals \cite{ISABC_TAP, WiFi_imaging_TAP, Vincent_Liu_thin_air}. By eliminating the need for power-intensive active RF components, this paradigm significantly improves energy efficiency while reducing hardware complexity and cost. Passive and semi-passive backscatter tags typically operate in the nanowatt-to-microwatt power range, making them highly suitable for large-scale and sustainable Internet-of-Things (IoT) deployments \cite{Van_AmBC_Survey}. Owing to their ultra-low power operation, low maintenance requirements, and flexible deployment, BC technologies have been widely adopted in applications such as environmental monitoring, precision agriculture, and smart city infrastructure \cite{Sensable_city}.

Importantly, in many smart environments, sensing plays a critical role alongside communication. In this context, ambient IoT tags can serve dual functionalities as both sensing entities and communication nodes. By analyzing the reflected RF signals from these tags, a receiver can extract spatial parameters for localization, tracking, and environmental awareness while simultaneously decoding the embedded information. This tight integration of passive communication and sensing paves the way for intelligent and energy-efficient wireless ecosystems \cite{Imaging_localization_backscatter}.

\begin{table*}[t]
\centering
\caption{{Contrasting our contributions with existing ISABC literature}}
\label{tab:lit_rev}
\small
\begin{tabular}{|l|c|c|c|c|c|c|}
\hline
\rowcolor{black!15}
\textbf{Features} &
\cite{Radar_enabled} &
\cite{ISBAC_letter} &
\cite{Beamforming_ABISAC} &
\cite{OFDM_delay_AoA_estimate} & \cite{AmBC_ISAC_V2X}
 &
\textbf{Proposed} \\
\hline

Targets & $\times$ & Active & Active & Active & Active  & Passive \\ \hline
\rowcolor{blue!15}
Clutters & \checkmark & $\times$ & $\times$ & $\times$ &  $\times$  & \checkmark \\ \hline
Tag's carrier waveform & Radar & Comm & Comm & Comm & Comm & Radar \\ \hline
\rowcolor{blue!15}Delay estimation & $\times$ & $\times$ & $\times$ & \checkmark & \checkmark  & \checkmark \\ \hline
Doppler estimation & $\times$ & $\times$ & $\times$ & \checkmark & \checkmark & \checkmark \\ \hline
\rowcolor{blue!15}Tag data-decoding & \checkmark & \checkmark & $\times$ & \checkmark & \checkmark  & \checkmark \\ \hline

\end{tabular}
\end{table*}
\subsection{Related works}
Towards channel estimation in AmBC systems, conventional minimum mean square error (MMSE) estimation is generally impractical due to the lack of accurate channel covariance information and the unknown ambient RF source. To address this challenge, Ma \textit{et al.} \cite{Ma_Shuo_blind_AmBC} proposed a blind expectation maximization (EM)-based channel estimation algorithm to estimate the magnitudes of the channel coefficients without requiring pilot signals. However, the absence of RF source knowledge limits its estimation accuracy. To further improve estimation performance, pilot-assisted channel estimation techniques have been investigated. In particular, Zhao \textit{et al.} \cite{Zhao_Channel_estimation_pilot} proposed a two-stage framework for AmBC systems with a massive-antenna reader, which jointly estimates the channel coefficients and directions of arrival (DoAs), followed by angular refinement. Furthermore, the authors of \cite{Channel_estimation_fullduplex} propose and compare three distinct solutions offering different accuracy-complexity tradeoffs: a pilot-based maximum-likelihood (ML) estimator and two semi-blind estimators based on EM and decision-directed (DD) frameworks for a full-duplex AmBC system. Nevertheless, these existing works are primarily designed for channel estimation to facilitate reliable tag data decoding and do not exploit the estimated channel parameters for integrated sensing and communication (ISAC) functionalities.

Motivated by the complementary capabilities of AmBC and ISAC, recent studies have investigated their integration to simultaneously support low-power communication and environmental sensing. In \cite{Radar_enabled}, radar clutter was exploited as a periodic carrier to enable ambient backscatter communication through dedicated tag encoding and decoding schemes. The integrated sensing and ambient backscatter communication (ISABC) framework proposed in \cite{ISBAC_letter} enables a backscatter tag to simultaneously function as a sensing target and an information transmitter, while a joint power allocation strategy enhances both sensing and communication performance. Leveraging orthogonal frequency-division multiplexing (OFDM) waveforms, \cite{OFDM_delay_AoA_estimate} proposed a compressed sensing-based framework for joint tag localization and data decoding. In the context of vehicle-to-everything (V2X) networks, \cite{AmBC_ISAC_V2X} introduced an ambient backscatter-assisted ISAC architecture, where BD mounted on targets reflect ISAC signals to enable joint target parameter estimation and data recovery. Furthermore, \cite{Beamforming_ABISAC} investigated a joint beamforming design to minimize the transmit power while satisfying the quality-of-service requirements of backscatter communication.

The main contributions of this work are as follows:
\begin{itemize}
    \item Distinct from existing works \cite{ISBAC_letter, OFDM_delay_AoA_estimate, AmBC_ISAC_V2X}, which assume that the tag is co-located with the target and actively participates in joint sensing and communication, we propose a novel AmBC-enabled ISAC framework in which the backscatter tag operates independently of the target and passively modulates its data over radar echoes reflected from both the target and the surrounding clutter.

    \item A comprehensive end-to-end signal model is derived, revealing the highly coupled nature of the estimation problem. To address this challenge, an efficient estimation methodology is developed to jointly estimate the target and clutter parameters while decoding the tag message.

    \item Extensive simulation results validate the effectiveness of the proposed framework. The Cramér–Rao lower bound (CRLB) of target parameters is analytically derived, and estimation performance is benchmarked against it. Furthermore, the inherent trade-off between sensing accuracy and communication performance in integrated sensing and ambient backscatter communication (ISABC) systems is comprehensively examined.
\end{itemize}

\textbf{Notations}: Throughout this paper, vectors are denoted by bold lowercase letters and matrices by bold uppercase letters. The symbols $\mathbf{A}(m,:)$ and $\mathbf{A}(:,n)$ indicate the $m$th row and $n$th column of a matrix $\mathbf{A}$, respectively.  The operators $(\cdot)^T$, $(\cdot)^H$, $(\cdot)^*$, $(\cdot)^\dagger$, and $\mathrm{Tr}(\cdot)$ stand for the transpose, Hermitian transpose, complex conjugate, Moore–Penrose pseudoinverse, and trace, respectively.  The operator $\mathrm{diag}(\cdot)$ forms a vector from the diagonal entries of its matrix argument. The products $\odot$ and $\circ$ denote the Hadamard and Khatri–Rao products, respectively. The operator $\mathrm{vec}(\cdot)$ stacks the columns of a matrix into a single column vector (column-major order).

\section{System Model }
As illustrated in Fig.~1, we consider a system in which a single-antenna radar transmitter illuminates a scattering scene using probing pulses. A bistatically positioned single-antenna dual-function radar-communication (DFRC) receiver collects the reflected radar echoes to estimate the 
target parameters. In addition, a passive backscatter tag located near the DFRC receiver leverages the incident radar waveform, as well as the clutter reflections from surrounding environment as carriers to modulate its information bits. Consequently, the DFRC receiver 
observes a superposition of the radar echoes and the tag-modulated backscatter signals. The receiver, therefore, performs joint processing to estimate the target parameters while simultaneously decoding the tag's message.

\textcolor{black}{The BD/tag conveys information by dynamically tuning the load impedance $  Z_L  $ connected to its antenna, a technique known as load modulation \cite{Van_AmBC_Survey,Kaplan_TWC}. This controls the complex reflection coefficient $  G  $, expressed as$$G = \frac{Z_L - Z_a^*}{Z_L + Z_a},$$where $  Z_a  $ denotes the antenna impedance.Depending on the selected load state, the BD operates in one of the following modes:
\begin{itemize}
\item \textit{Open-circuit state} ($  G \approx +1  $): When $  Z_L \rightarrow \infty  $, the incident signal is almost entirely reflected with negligible phase shift, resulting in maximum backscattered power.
\item \textit{Short-circuit state} ($  G \approx -1  $): When $  Z_L = 0  $, the incident signal is nearly fully reflected with a $180^\circ$ phase shift, enabling binary phase modulation.
\item \textit{Matched (absorption) state} ($  G \approx 0  $): When $  Z_L = Z_a^*  $, most of the incident energy is absorbed with minimal reflection. This state can also be utilized for energy harvesting at the BD.
\end{itemize}In practice, the BD switches among these discrete impedance states using a low-power RF switch or tunable load network. The combination of open- and short-circuit states ($  G \approx \pm 1  $) realizes binary phase-shift keying (BPSK), while switching between reflective and absorptive states ($  G \approx 0/1  $) implements on-off keying (OOK). Due to hardware non-idealities and impedance mismatches, the magnitude of the reflection coefficient typically satisfies $  |G| < 1  $ in real implementations. Thanks to this simple modulation scheme, the encoding complexity at the tag remains extremely low. At the DFRC receiver, the decoding effort associated with such backscatter signaling is minimal compared to the receiver and does not meaningfully affect the overall system complexity. }

{\color{black}
\textbf{Operating Conditions:}
\begin{itemize}

\item \textit{Quasi-static reflection coefficients (Swerling-I model):}
The target reflection coefficients follow the Swerling-I model, wherein the radar cross section (RCS) remains constant over a coherent processing interval (CPI) and varies independently across different realizations.

\item Over the CPI, the tag absorption coefficient $G$ and pilot power remain constant.

\item \textit{Unambiguous delay--Doppler region:}
The delay--Doppler pairs $(\tau_\ell,\nu_\ell)$ lie within the unambiguous region, i.e., $\tau_\ell \in [0,\tau_{\max}]$ and $\nu_\ell \in [-\nu_{\max},\nu_{\max}]$, where $\tau_{\max} = 1/\Delta f$ and $\nu_{\max} = 1/(2T)$, ensuring alias-free estimation.

\item \textit{Slowly varying delays:}
The propagation delays are approximately constant within a CPI, as the variation due to target motion is negligible compared to the delay resolution.

\item \textit{Constant Doppler shifts:}
The Doppler frequencies remain constant over the CPI, which is valid when the target velocity variation is sufficiently small.

\end{itemize}
}

\subsection{Transmit Signals}
Consider a pulse-Doppler radar that transmits a train of \( P \) pulses. The total duration of these \( P \) pulses is referred to as the coherent processing interval (CPI). The radar transmits a baseband pulse signal \( s(t) \) periodically, with a pulse repetition interval (PRI) of \( T \), such that the transmitted signal is given by
\begin{align}
    x_r(t) =\sum_{p=0}^{P-1}s(t - pT), 0 < t < (P - 1)T,
    \label{Radar_Pulse_train}
\end{align}
where the pulse $s(t)$ is a time-limited baseband function, whose continuous-time Fourier transform is $S(f) = \int_{-\infty}^{\infty} s(t)e^{-j2\pi f t}dt$. In general, most of the radar signal’s energy lies within the spectrum $[-B/2, B/2]$, where $B$ denotes the effective signal bandwidth. \par
The DFRC receiver captures radar echoes from targets arriving directly and via the backscatter tag. The tag modulates the incident waveform, comprising $L$ target reflections and $Q$ clutter interference, by switching its antenna load between distinct impedance states to encode the message \( g(t) \).

\begin{figure*}[t]
\centering
\includegraphics[width=0.75\textwidth]{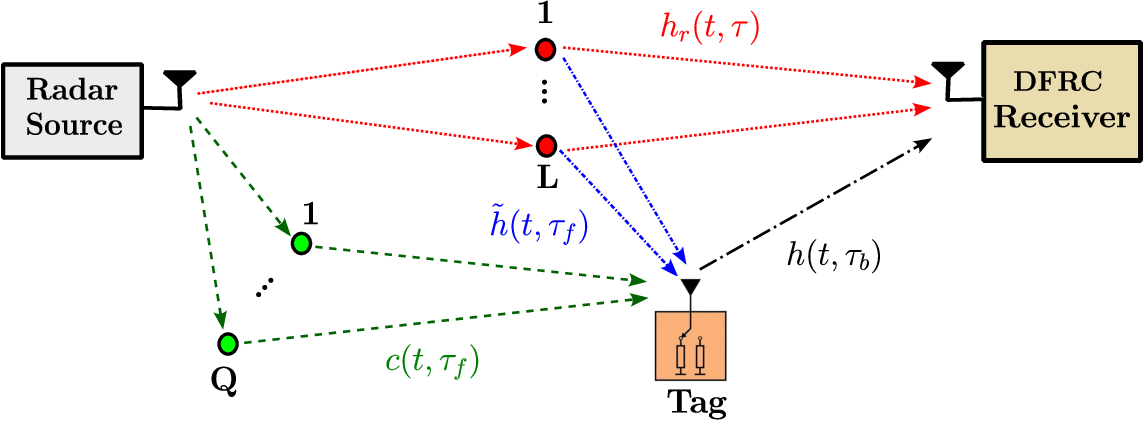}
\caption{\textcolor{black}{Illustration of the ISABC system model comprising a radar source, $L$ targets, and $Q$ clutter scatterers. The red dotted lines denote target reflection paths, while the green dotted lines represent clutter-induced multipath components. The blue dashed path corresponds to the target-to-tag backscatter link, where the tag modulates the incident signal and forwards it to the DFRC receiver via a direct LoS connection, along with other propagation paths. }}
\label{fig:AmBC_JRC_System_Model}
\end{figure*}

\subsection{Channels}
\subsubsection{Radar channel between source and DFRC receiver}
The target scene comprises $L$ non-fluctuating point targets, according to the Swerling-I target model. The unknown target parameter vectors are 
${\boldsymbol{\alpha}_r} \in \mathbb{C}^L$, ${\boldsymbol{\bar{\tau}}}_r \in \mathbb{R}^L$, ${\boldsymbol{\bar{\nu}}}_r \in \mathbb{R}^L$, and where the $\ell$th target is characterized by: time delay $[\boldsymbol{\bar{\tau}_r}]_{\ell}$, which is linearly proportional to the target’s range i.e. $[\boldsymbol{\bar{\tau}}_r]_{\ell}= d_{\ell}/c$ where $c$ is the speed of light and $d_{\ell}$ is the target bistatic range; Doppler frequency $[\boldsymbol{\bar{\nu}_r}]_{\ell}$, proportional to the target’s radial velocity i.e., $[\boldsymbol{\bar{\nu}_r}]_{\ell} = 4\pi f_c v_\ell/c$ where $v_\ell$ is the radial velocity and complex amplitude $[\boldsymbol{\alpha}_r]_{\ell}$
that models the path loss and reflectivity
of the $\ell$th target. The target's range and Doppler shift are assumed to lie in the unambiguous time-frequency region, i.e., the time delays are no longer than the PRI, and Doppler frequencies are up
to the pulse repetition frequency (PRF). The delay–Doppler radar channel impulse response can be expressed as a product of Dirac delta functions as \cite{DBD, Multi_DBD}
\begin{align}
h_r(\tau, \nu)=\sum_{\ell=1}^{L}\left[\boldsymbol{\alpha}_r\right]_{\ell} \delta\left(\tau-\left[{\boldsymbol{\bar{\tau}}}_r\right]_{\ell}\right) \delta\left(\nu-\left[\boldsymbol{\bar{\nu}}_r\right]_{\ell}\right).
\end{align}
Therefore, the time-delay representation of the channel is obtained by taking the Fourier transform along the Doppler axis as
\begin{align}
h_r(t, \tau)=\sum_{\ell=1}^{L}\left[\boldsymbol{\alpha}_r\right]_{\ell} \delta\left(\tau-\left[\overline{\boldsymbol{\tau}}_r\right]_{\ell}\right) e^{-\mathrm{j} 2 \pi\left[\overline{\boldsymbol{\nu}}_r\right]_{\ell} t}.
\label{Direct chanel}
\end{align}

\subsubsection{Channel between radar source and tag through $Q$ clutters}
The clutter channel comprises $Q$ propagation paths characterized by their gain, time-delay, Doppler shift respectively, encapsulated in the parameter vectors $\boldsymbol{\zeta}_c \in \mathbb{C}^Q$, ${\boldsymbol{\bar{\tau}}}_c \in \mathbb{R}^Q$, ${\boldsymbol{\bar{\nu}}}_c \in \mathbb{R}^Q$, respectively. 
Similar to the radar channel, the time-delay domain representation of the clutter channel is
\begin{align}
 c(t, \tau_f)=\sum_{q=1}^{Q}\left[\boldsymbol{\zeta}_{c}\right]_{q} \delta\left(\tau_f-\left[\overline{\boldsymbol{\tau}}_c\right]_{q}\right) e^{-\mathrm{j} 2 \pi\left[\overline{\boldsymbol{\nu}}_c\right]_{q} t}. 
  \label{Forward Channel between Radar Source and Tag through $Q$ multipath}
\end{align}

\subsubsection{ Channel between radar source and tag through $L$ targets}
The interference channel comprises $L$ targets characterized by their gain, time-delay, Doppler shift respectively, encapsulated in the parameter vectors $\boldsymbol{\zeta}_i \in \mathbb{C}^L$, ${\boldsymbol{\bar{\tau}}}_i \in \mathbb{R}^L$, ${\boldsymbol{\bar{\nu}}}_i \in \mathbb{R}^L$. 
The time-delay representation of the channel is 
\begin{align}
   \tilde{h}(t, \tau_f)=\sum_{\ell=1}^{L}\left[\boldsymbol{\zeta}_{i}\right]_{\ell} \delta\left(\tau_f-\left[\overline{\boldsymbol{\tau}}_i\right]_{\ell}\right) e^{-\mathrm{j} 2 \pi\left[\overline{\boldsymbol{\nu}}_i\right]_{\ell} t}.
   \label{Interference Channel between Radar Source and Tag through L targets}
\end{align}

\subsubsection{Backscattered channel between tag and DFRC receiver}
The backscattered channel comprises line-of-sight (LoS) path characterized by their gain and time-delay in the parameter ${{\zeta}_{b}} \in \mathbb{C}$ and ${{\bar{\tau}}}_b \in \mathbb{R}$ is expressed as
\begin{align}
   {h}(t, \tau_b)=\left[{\zeta}_b\right] \delta\left(\tau_b-\left[\overline{{\tau}}_b\right]\right). 
   \label{Backscattered Channel between Tag and DFRC receiver}
\end{align}
\textcolor{black}{The propagation delay between the tag and the DFRC receiver is treated as a known constant, since the tag is assumed to be stationary with a known \textit{a priori} location relative to the receiver. When explicitly modeled, this delay acts as an additive, known constant in the cascaded (tag-assisted) propagation path.}

\subsection{Received Signal at DFRC receiver}
\subsubsection{Received signal through $L$ targets} The direct received signal \( y_r(t) \), resulting from the propagation between the source and the DFRC receiver via reflections from \( L \) distinct targets is given by
\begin{align}
 {y}_r(t) ={h}_r(t) * x_r(t)
=\int_{-\infty}^{\infty}h_r(t, \tau) x_r(t-\tau) d \tau .    
\label{Rx signal direct}
\end{align}
Substituting the expressions of the radar channels from \eqref{Direct chanel} and the transmit signal $x_r(t)$, we obtain
\begin{equation}
 \begin{aligned}
{y_r(t) }  = \sum_{p=0}^{P-1}\sum_{\ell=1}^{L}\left[\boldsymbol{\alpha}_r\right]_{\ell} & s\left(t-p T-\left[\overline{\boldsymbol{\tau}}_r\right]_{\ell}\right) e^{-\mathrm{j} 2 \pi\left[\overline{\boldsymbol{\nu}}_r\right]_{\ell} t}.
\end{aligned}   
\end{equation}

\subsubsection{Received Signal at DFRC receiver through $Q$ clutter via tag}
 The received signal through $Q$ clutters via tag is \cite{Griffin_RF_tags_MIMO}, \cite{Van_AmBC_Survey}
 \begin{equation}
   \begin{aligned} 
   y_{ct}(t)=  \int_{-\infty}^{+\infty} \int_{-\infty}^{+\infty}  &{h}_{b}(t, \tau_b) {g}(t) c(t, \tau_f) \\
   & \times {x}_r\left(t-\tau_b-\tau_f\right) d \tau_b d \tau_f,
   \end{aligned}  
 \end{equation}   
where $g(t)$ is the tag information. Substituting the expressions of the radar channels from \eqref{Forward Channel between Radar Source and Tag through $Q$ multipath}, \eqref{Backscattered Channel between Tag and DFRC receiver} and the transmit signal $x_r(t)$, we obtain
\begin{equation}
    \begin{aligned}
   y_{ct}(t)& = \int_{-\infty}^{+\infty} \int_{-\infty}^{+\infty} \left[{\zeta_b}\right] \delta\left(\tau_b\right)g(t) \\
   &\times \sum_{q=1}^{Q}\left[\boldsymbol{\zeta}_{c}\right]_{q} \delta\left(\tau_f-\left[\overline{\boldsymbol{\tau}}_c\right]_{q}\right)
    e^{-\mathrm{j} 2 \pi\left[\overline{\boldsymbol{\nu}}_c\right]_{q} t} \\
    & \times {x}_r\left(t-\tau_b-\tau_f\right) d \tau_b d \tau_f,
   \end{aligned}
\end{equation}
which is further expressed as
\begin{equation}
    \begin{aligned}
   y_{ct}(t)=\sum_{q=1}^{Q} \left[\boldsymbol{\zeta_c}\right]_{q}\left[{\zeta}_b\right]&{x}_r\left(t-\left[\overline{\boldsymbol{\tau}}_c\right]_{q}\right)
   e^{-\mathrm{j} 2 \pi\left[\overline{\boldsymbol{\nu}}_c\right]_{q} t}g(t).
\end{aligned}
\end{equation}
Let $ \left[\boldsymbol{\zeta}_{c}\right]_{q}\left[{\zeta}_b\right]=\left[\boldsymbol{\gamma}_c\right]_{q}$ and substitute the value of the transmit signal $x_r(t)$, we obtain
\begin{equation}
  \begin{aligned}
    y_{ct}(t)= \sum_{p=0}^{P-1}\sum_{q=1}^{Q}\left[\boldsymbol{\gamma}_c\right]_{q} & s\left(t-pT-\left[\overline{\boldsymbol{\tau}}_c\right]_{q}\right)e^{-\mathrm{j} 2 \pi\left[\overline{\boldsymbol{\nu}}_c\right]_{q} t} g(t).
\end{aligned}  
\end{equation}

\subsubsection{Received Signal at DFRC receiver through $L$ targets via tag}
 The received signal corresponding to the $L$ targets via the tag is similarly given by
 \begin{equation}
     \begin{aligned}
   y_{t t}(t)= \sum_{p=0}^{P-1}\sum_{\ell=1}^{L}\left[\boldsymbol{\psi}_i\right]_{\ell}& s\left(t-pT-\left[\overline{\boldsymbol{\tau}}_i\right]_{\ell}\right)e^{-\mathrm{j} 2 \pi\left[\overline{\boldsymbol{\nu}}_i\right]_{\ell} t} g(t),
\end{aligned} 
 \end{equation} 
where  $ \left[\boldsymbol{\psi}_c\right]_{\ell}= \left[\boldsymbol{\zeta}_i\right]_{\ell}\left[{\zeta}_b\right]$.\par
Therefore, the received signal at the DFRC receiver is the superposition of ${y_r(t) }$, $y_{ct}(t)$ and $ y_{tt}(t)$
which is expressed as
\begin{equation}
    \begin{aligned}
 y(t)=& \sum_{p=0}^{P-1}\Bigg(\sum_{\ell=1}^{L}\left[\boldsymbol{\alpha}_r\right]_{\ell} s\left(t-p T-\left[\overline{\boldsymbol{\tau}}_r\right]_{\ell}\right) e^{-\mathrm{j} 2 \pi\left[\overline{\boldsymbol{\nu}}_r\right]_{\ell} pT}   \\
 &+\sum_{q=1}^{Q}\left[\boldsymbol{\gamma}_c\right]_{q} s\left(t-pT-\left[\overline{\boldsymbol{\tau}}_c\right]_{q}\right)e^{-\mathrm{j} 2 \pi\left[\overline{\boldsymbol{\nu}}_c\right]_{q} pT} g(t)\\
 &+\sum_{\ell=1}^{L}\left[\boldsymbol{\psi}_i\right]_{\ell} s\left(t-pT-\left[\overline{\boldsymbol{\tau}}_i\right]_{q}\right)e^{-\mathrm{j} 2 \pi\left[\overline{\boldsymbol{\nu}}_i\right]_{q} pT} g(t) \Bigg)  ,
 \label{final_system_model}
\end{aligned}
\end{equation}
where the last approximation follows from $\left[\overline{\boldsymbol{\nu}}_r\right]_{\ell} pT \ll 1$ , $\left[\overline{\boldsymbol{\nu}}_c\right]_{q} pT \ll 1$ and $\left[\overline{\boldsymbol{\nu}}_i\right]_{q} pT \ll 1$ \cite{Zheng_Super_resolution}, which results in a
constant phase rotation within one coherence period.\\

\begin{figure*}[!t]
\centering
\includegraphics[width=0.9\textwidth]{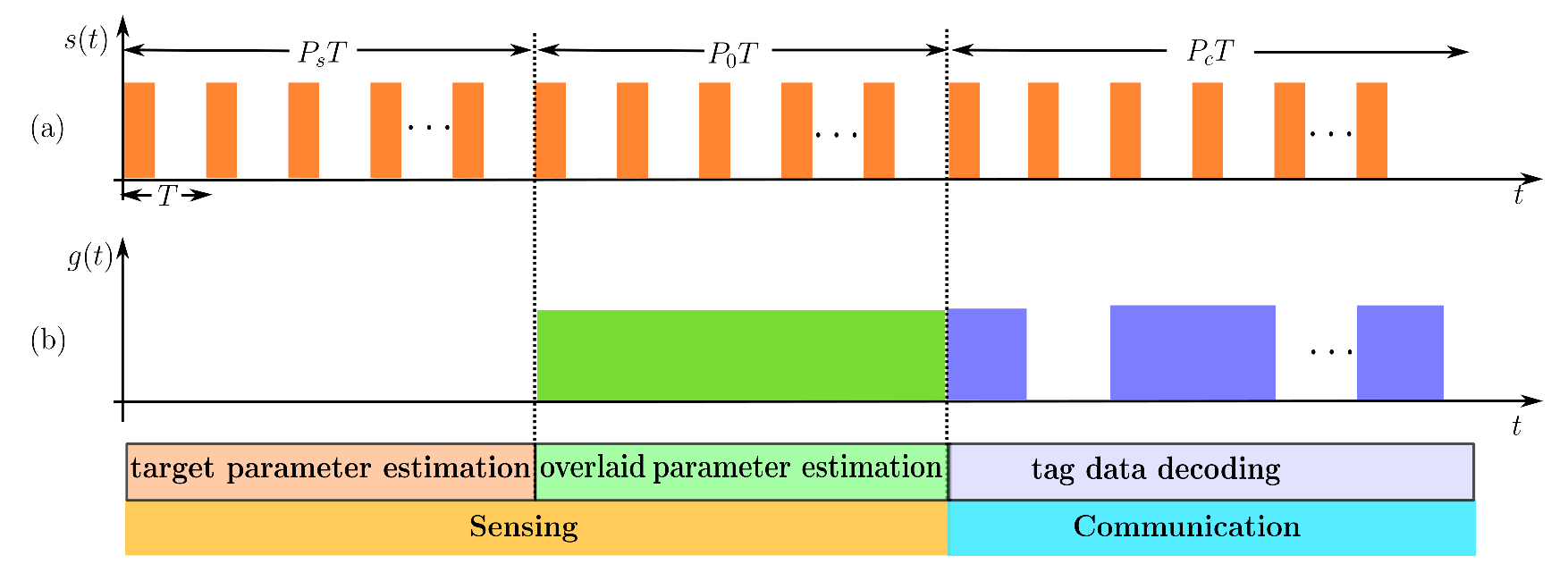}
\caption{(a) Radar source transmitting a total of $P = P_s + P_0 + P_c$ pulses; (b) Tag operation phases, including energy harvesting during $P_s$, pilot transmission during $P_0$, and data transmission during $P_c$.}
\label{fig:tx_signal}
\end{figure*}

\section{Parameter Estimation}
The main objective of this paper is to estimate the target parameters, clutter parameters, and decode the tag message from the superimposed observation $y(t)$, where the total number of unknown parameters is $(2L + Q)$.
To achieve this, we propose a novel frame structure that effectively facilitates these estimations. Toward this, the $P$ radar pulses are divided into three parts, as shown in Fig. 2:\\
\textbf{Stage-I:} During the interval \( 0 < t < (P_s - 1)T \), the tag operates in the absorption state i.e. $g(t) =0$, \textcolor{black}{energy harvesting mode} which vanishes the second and third term of \eqref{final_system_model}. This reduces the effective system model to the conventional radar system model.\\
\textbf{Stage-II:} During $(P_s-1)T<t<(P_s + P_0-1)T$ the tag transmit a known pilot signal $g_p(t) = \sqrt{G}$, where $G$ is the power budget at the tag. By incorporating the knowledge of the estimated parameters in step (i), only the second and third terms will be available. This leads to an overlaid channel estimation problem.\\
\textbf{Stage-III}: During $(P_s + P_0-1)T<t<(P-1)T$, the tag transmits the unknown data symbols, which can be decoded by utilizing the knowledge of other estimated parameters in steps (i) and (ii).\\
\textit{Remark}: The relation $P_s + P_0 + P_c = P$ characterizes the inherent trade-off between the sensing 
and communication functionalities, since these operations orthogonally partition the available 
pulse resources. Increasing $P_s$ allocates more pulses to sensing, which enhances the accuracy 
of delay--Doppler estimation due to the availability of more observational samples. However, 
this reduces the number of pulses available for tag communication, thereby lowering the 
achievable data rate. Conversely, decreasing $P_s$ dedicates a larger fraction of pulses to 
backscatter communication, improving the tag's throughput, but at the expense of degraded 
sensing performance resulting from fewer sensing snapshots. This interplay highlights the 
fundamental balance between reliable target sensing and efficient tag communication in the 
considered ISABC system.\par
Although, in principle, one might attempt to estimate both the target parameters and the tag data using only a two-stage approach—by applying the second stage to jointly estimate all target and clutter parameters—this strategy proves inadequate in practice. Specifically, when the second stage is applied directly, the same physical targets appear twice in the delay–Doppler domain at different locations. This occurs because the echoes received directly from the targets and those arriving via the tag-assisted path are associated with different effective delays, Doppler shifts, and gains. As a result, even though all target and clutter parameters can be estimated, it becomes impossible to discriminate between the parameters corresponding to the direct path and those corresponding to the tag-reflected path of the same targets. This fundamental ambiguity motivates adopting the proposed three-stage procedure rather than a two-stage one. The resulting ambiguity is explicitly demonstrated in Section IV via the delay–Doppler imaging results.

\subsection{Parameter estimation of $L$ targets}
During the first $P_s$ pulses, the tag remains silent with $g(t)=0$. Accordingly, the system model in \eqref{final_system_model} reduces to
\begin{equation}
 \begin{aligned}
{y(t) } & =\sum_{p=0}^{P_s-1}\sum_{\ell=1}^{L}\left[\boldsymbol{\alpha}_r\right]_{\ell} s\left(t-p T-\left[\overline{\boldsymbol{\tau}}_r\right]_{\ell}\right) e^{-\mathrm{j} 2 \pi\left[\overline{\boldsymbol{\nu}}_r\right]_{\ell} p T}  \\
& = \sum_{p=1}^{P-1}\widetilde{y}_p(t) ,
\label{ypt}
\end{aligned}   
\end{equation}
where $\widetilde{y}_p(t)$ is the observation corresponding to the $p$th pulse. Defining the time-aligned signals as $y_p(t)=\widetilde{y}_p(t+pT)$ ensures that $y_p(t)$, for $p=0,\ldots,P_s-1$, are synchronized with $y_0(t)$. The CTFT of $y_p(t)$ is given by
\begin{equation}
    \begin{aligned}
    \label{CTFT}
  {Y}_p(f)&=\int_{p T}^{p T+T}y_p(t)  e^{-\mathrm{j} 2 \pi f t} dt\\
        &=
\sum_{\ell=1}^{L}\left[\boldsymbol{\alpha}_r\right]_{\ell} {S}(f) e^{-\mathrm{j} 2 \pi\left(\left[\overline{\boldsymbol{\tau}}_r\right]_{\ell} f+\left[\overline{\boldsymbol{\nu}}_r\right]_{\ell} pT\right)}.
    \end{aligned}
\end{equation}
After sampling $ {Y}_p(f)$ uniformly at the frequencies $f_n = n \Delta f = n B/M$  ,
with $n = 0,...,N-1$, yield
\begin{equation}
\begin{aligned}
&Y_p(f_n) =
\sum_{\ell=1}^{L} \left[\boldsymbol{\alpha}_r\right]_{\ell} 
e^{-j 2\pi \left( \left[\overline{\boldsymbol{\tau}}_r\right]_{\ell} n\Delta f + \left[\overline{\boldsymbol{\nu}}_r\right]_{\ell} pT \right)} S(f_n) \\
&\Rightarrow Y_p(f_n) = \left[ Y \right]_{p,n} =
\sum_{\ell=1}^{L} \left[\boldsymbol{\alpha}_r\right]_{\ell} 
e^{-j 2\pi \left( n \left[ \boldsymbol{\tau}_r \right]_{\ell} + p \left[ \boldsymbol{\nu}_r \right]_{\ell} \right)} S(f_n).
\end{aligned}
\label{Ypn_combined}
\end{equation}
where $\boldsymbol{\tau}_r = \overline{\boldsymbol{\tau}}_r \Delta f$
 is the normalized delay, $\boldsymbol{\nu}_r = \overline{\boldsymbol{\nu}}_r/T
$ is the normalized Doppler frequency with ${\boldsymbol{\tau}}_r \in [0,1]$ and  ${\boldsymbol{\nu}}_r \in [-0.5,0.5]$.
  
By arranging the received samples $\{Y_{p,n}\}$, with $P_s$ temporal snapshots indexed along the rows and $N$ frequency-domain samples indexed along the columns, the observation matrix $\mathbf{Y}_r \in \mathbb{C}^{P_s \times N}$ is formed as  
\begin{equation}
    \begin{aligned}
    \mathbf{Y}_r  = \mathbf{S} \odot (\mathbf{D}(\boldsymbol{\nu}_r) \mathrm{diag}(\boldsymbol{\alpha}_r) \mathbf{C}^T(\boldsymbol{\tau}_r) ) + \mathbf{E}_r
    \end{aligned}
    \label{eq:2D_MUSIC_system_model}
\end{equation}
where $\mathbf{D}(\boldsymbol{\nu}_r) \in \mathbb{C}^{P_s \times L}$ is the Doppler manifold matrix, whose $(p,\ell)$th element is given by 
$e^{-\mathrm{j} 2\pi (p-1) [\boldsymbol{\nu}_r]_{\ell}}$, and $\mathbf{C}(\boldsymbol{\tau}_r) \in \mathbb{C}^{N \times L}$ is the delay manifold matrix, whose $(n,\ell)$th element is given by 
$e^{-\mathrm{j} 2\pi (n-1) [\boldsymbol{\tau}_r]_{\ell}}$, respectively. Furthermore, the matrix $\mathbf{S}= \mathbf{1}_{P_s}\mathbf{s} \in \mathbb{C}^{P_s \times N}$, where the vector $\mathbf{s} = [S(f_0), S(f_1), \ldots, S(f_{N-1})]^T$. Additionally, the matrix $\mathbf{E}_r$ denotes additive white Gaussian noise (AWGN).
  \par 
To recover the target parameters, the two-dimensional multiple signal classification (2D-MUSIC) 
algorithm is employed to enable high-resolution joint estimation of the delay and Doppler frequencies. Since the MUSIC algorithm requires multiple snapshots of the received waveform, spatial smoothing is applied beforehand to ensure reliable estimation. Specifically, the steering vector used in the 2D-MUSIC algorithm is defined as \textcolor{black}{\cite{Zheng_Super_resolution}}:
\begin{align}
 \mathbf{a}(\nu, \tau) = \mathrm{vec}(\mathbf{A}(\nu, \tau)) \in \mathbb{C}^{P_s N \times 1}   
\end{align}
where the $(p',n')$th element of 2D steering matrix $\mathbf{A}(\nu, \tau)$ 
is $e^{\mathrm{j} 2\pi (-(p'-1)\nu - (n'-1)\tau)}$ where $p' = 1,2,\hdots,P'<P_s$ and $n' = 1,2,\hdots,N'<N$.
Consider $N_{\text{s}} = (N - N' + 1)(P_s - P' + 1)$ snapshots after spatial smoothing, grouping the received signals and vectorizing each matrix yields the observation matrix $\mathbf{Z} \in \mathbb{C}^{P'N' \times N_{\text{s}}}$ as
\begin{align}
 \mathbf{Z} = \begin{bmatrix}
\mathrm{vec}(\breve{\mathbf{Y}}_{0,0}), .. , \mathrm{vec}(\breve{\mathbf{Y}}_{N - N', 0}),.. , \mathrm{vec}(\breve{\mathbf{Y}}_{N - N', P_s - P'})
\end{bmatrix}.   
\end{align}
Let $\Bar{\mathbf{Y}}_{p',n'} = \frac{\mathbf{Y}_r(p',n')}{\mathbf{S}(p',n')} \forall p=0,1,\hdots,P_s - P'$ and $n=0,1,\hdots,N-N'$, then  $\breve{\mathbf{Y}}_{p',n'}$ is defined as \textcolor{black}{\cite{Zheng_Super_resolution}}
\begin{equation}
\breve{\mathbf{Y}}_{p',n'} =
\begin{bmatrix}
\Bar{\mathbf{Y}}_{p',n'} & \cdots & \Bar{\mathbf{Y}}_{p',n'+N'-1} \\
\vdots & \ddots & \vdots \\
\Bar{\mathbf{Y}}_{p'+P'-1,n'} & \cdots & \Bar{\mathbf{Y}}_{p'+P'-1,n'+N'-1}
\end{bmatrix}
\end{equation}

After applying spatial smoothing, the observation matrix \(\mathbf{Z}\) is subjected to singular value decomposition (SVD) as $\mathbf{Z} = \mathbf{U} \mathbf{\Sigma} \mathbf{V}^{{H}}$,
where \(\mathbf{U} \in \mathbb{C}^{P' N' \times P'N'}\) and \(\mathbf{\Sigma} \in \mathbb{C}^{N_s \times N_s}\). The matrix \(\mathbf{U}\), containing the left singular vectors, is partitioned into the signal and noise subspaces as
$\mathbf{U} = [\mathbf{U}_{s}, \mathbf{U}_{n}]$, 
where \(\mathbf{U}_{s}\) and \(\mathbf{U}_{n}\) represent the signal and noise subspaces, respectively. In practice, the signal subspace is identified by selecting the eigenvectors corresponding to the dominant singular values.
Finally, the 2D-MUSIC spectrum is constructed by exploiting the orthogonality of the steering vector to the noise subspace, and is given by
\[
P_{\text{MUSIC}}(\nu, \tau) = \frac{1}{\mathbf{a}^H(\nu, \tau) \mathbf{U}_{n} \mathbf{U}^H_{n} \mathbf{a}(\nu, \tau)}.
\]

The 2D-MUSIC algorithm then estimates the target's Doppler frequency and delay by identifying the peaks in this spectrum. To reduce the quantization error introduced by the discrete delay–Doppler grid, the initial delay and Doppler estimates are subsequently refined using a golden-section search. Once the delay and Doppler frequency components are resolved, the corresponding amplitude coefficients are estimated using a least-squares (LS) method. Toward this, the delay and Doppler frequency array manifold $\mathbf{D}(\boldsymbol{\nu}_r)$ and $\mathbf{C}(\boldsymbol{\tau}_r)$ can be estimated using the estimated delay and Doppler. To facilitate a linear representation, the system model in \eqref{eq:2D_MUSIC_system_model} is expressed as
\begin{equation}
    \begin{aligned}
    \mathbf{Y}_r  =  \mathbf{D}(\boldsymbol{\nu}_r) \mathrm{diag}(\boldsymbol{\alpha}_r) \mathbf{C}^T(\boldsymbol{\tau}_r) \widetilde{\mathbf{S}} + \mathbf{E}_r.
    \end{aligned}
    \label{eq:2D_MUSIC_system_model2}
\end{equation}
where $\widetilde{\mathbf{S}} = \mathrm{diag}(\mathbf{s})$. Subsequently, vectorizing the observation yields $\mathbf{y}_r = \mathrm{vec}(\mathbf{Y}_r)$ as
\begin{equation}
    \begin{aligned}
    \mathbf{y}_r  = \boldsymbol{\Psi}\boldsymbol{\alpha}_r   + \mathbf{e}_r
    \end{aligned}
    \label{LS}
\end{equation}
where the matrix $\boldsymbol{\Psi}$ is defined as $\boldsymbol{\Psi} = \left( \widetilde{\mathbf{S}}^T\mathbf{C}(\boldsymbol{\tau}_r) \circ  \mathbf{D}(\boldsymbol{\nu}_r) \right)$. Finally, the least-squares (LS) estimate of $\boldsymbol{\alpha}_r$ is given by $\widehat{\boldsymbol{\alpha}}_r = \boldsymbol{\Psi}^{\dagger} \mathbf{y}_r$.

\textcolor{black}{After estimating the $L$ direct-path target parameters in Stage-I, the corresponding component in~(14) is reconstructed and subtracted from the received signal to obtain a residual observation. The resulting signal comprises the clutter-induced components and the indirect (tag-assisted) target reflections. Although the indirect component originates from the same set of targets, it undergoes an additional reflection via the tag and propagates over a longer bistatic path. Consequently, it exhibits distinct delay, Doppler, and complex gain parameters compared to the direct-path signal. This separation in the delay--Doppler domain ensures that the subtraction of the direct-path estimate does not distort the indirect component, which remains available for subsequent processing in the following stages.}

\subsection{Overlaid Parameter Estimation}
After estimating the $L$ target parameters, the first term from the \eqref{final_system_model} can be removed. Furthermore, by applying the procedure in \eqref{ypt}--\eqref{Ypn}, the overlaid system model is expressed as
\begin{equation}
    \begin{aligned}
 \left[{{Y}_c}\right]_{p,n}=&
\sum_{q=1}^{Q}\left[\boldsymbol{\gamma}_c\right]_{q}  e^{-\mathrm{j} 2 \pi\left(n\left[{\boldsymbol{\tau}}_c\right]_{q} +p\left[{\boldsymbol{\nu}}_c\right]_{q} \right)}{S}(f_n) \sqrt{G}\\&+\sum_{\ell=1}^{L}\left[\boldsymbol{\psi}_i\right]_{\ell}  e^{-\mathrm{j} 2 \pi\left(n\left[{\boldsymbol{\tau}}_i\right]_{\ell} +p\left[{\boldsymbol{\nu}}_i\right]_{\ell} \right)}{S}(f_n)\sqrt{G}.
\label{Ypn}
    \end{aligned}
\end{equation}
Stacking the observations $\left[ \mathbf{Y}_c \right]_{p,n}$ for $P_s \leq p \leq P_s + P_0 - 1$ along the rows and $N$ samples along the columns yields the observation matrix $\mathbf{Y}_c \in \mathbb{C}^{P_0 \times N}$ as 
\begin{equation}
    \begin{aligned}
    \mathbf{Y}_c  = \mathbf{S}_c\odot (\mathbf{D}(\boldsymbol{\nu}_c) \mathrm{diag}(\boldsymbol{\eta}) \mathbf{C}^T(\boldsymbol{\tau}_c) ) + \mathbf{E}_c
    \end{aligned}
    \label{eq:2D_MUSIC_system_model3}
\end{equation}
where $\mathbf{S}_c = \mathbf{1}_{P_0}\bar{\mathbf{s}} \in \mathbb{C}^{P_0 \times N}$, where the vector $\bar{\mathbf{s}} = \sqrt{G} [S(f_0), S(f_1), \ldots, S(f_{N-1})]^T$. However,  $\mathbf{D}(\boldsymbol{\nu}_c) = [\mathbf{D}(\boldsymbol{\nu}_q) \quad \mathbf{D}(\boldsymbol{\nu}_\ell)] \in \mathbb{C}^{P_0 \times (Q+L)}$, where $\mathbf{D}(\boldsymbol{\nu}_q)$  and $\mathbf{D}(\boldsymbol{\nu}_\ell)$ represent the Doppler manifolds of the  $Q$ clutter components and $L$ targets, respectively. The vector $\boldsymbol\eta =[\boldsymbol\gamma_c^T, \boldsymbol\psi_c^T]^T\in \mathbb{C}^{Q+L}$ consists the complex gains of $Q$ clutters and $L$ targets. The $p$th row of both is given by  $e^{-\mathrm{j} 2\pi (P_s + p - 1)\nu}$, evaluated at the corresponding Doppler frequency. A similar structure is also adapted in $\mathbf{C}(\boldsymbol{\tau}_c).$ Notably, the system model in \eqref{eq:2D_MUSIC_system_model3} is similar to \eqref{eq:2D_MUSIC_system_model}. Therefore, the 2D-MUSIC is again invoked to estimate the parameters of $Q$ clutter and $L$ targets. \par
At first glance, the individual parameters of clutter and targets may appear unnecessary for tag data decoding. However, due to the variation in Doppler-induced phase shifts across different pulse indices in the third stage, estimating the individual parameters becomes essential, rather than relying on the aggregated channel response.

\subsection{Tag data decoding}
In the third stage, the tag transmits data in each pulse duration, i.e., $g(t)=g(p)$ for $P_s + P_0 \leq p \leq P-1$. Applying the procedure in \eqref{ypt}--\eqref{Ypn} yields the system model as
\begin{equation}
    \begin{aligned}
 \left[{{Y}_d}\right]_{p,n}=&
\sum_{q=1}^{Q}\left[\boldsymbol{\gamma}_c\right]_{q}  e^{-\mathrm{j} 2 \pi\left(n\left[{\boldsymbol{\tau}}_c\right]_{q} +p\left[{\boldsymbol{\nu}}_c\right]_{q} \right)}{S}(f_n) g(p) \\&+\sum_{\ell=1}^{L}\left[\boldsymbol{\psi}_i\right]_{\ell}  e^{-\mathrm{j} 2 \pi\left(n\left[{\boldsymbol{\tau}}_i\right]_{\ell} +p\left[{\boldsymbol{\nu}}_i\right]_{\ell} \right)}{S}(f_n)g(p).
    \end{aligned}
\end{equation}
Note that all target and clutter parameters are estimated in the previous stages. Given these parameters and the known training signal  $S(f_n)$, the linear model can now be formulated by arranging the observations $\left[{Y}_d\right]_{p,n}$, for indices $P_s + P_0 \leq p \leq P - 1$, along the rows, and subcarriers $n = 0, \ldots, N - 1$ along the columns. This results in the observation matrix $\mathbf{Y}_d \in \mathbb{C}^{P_c \times N}$, defined as
\begin{equation}
    \begin{aligned}
    \mathbf{Y}_d  = \mathbf{G}_d \mathbf{X} + \mathbf{E}_d,
    \end{aligned}
\end{equation}
\textcolor{black}{where $\mathbf{X} = \widetilde{\mathbf{D}}(\boldsymbol{\nu}_c)\,\mathrm{diag}(\boldsymbol{\eta})\,\mathbf{C}^T(\boldsymbol{\tau}_c)\,\widetilde{\mathbf{S}} \in \mathbb{C}^{P_c \times N}$. Since the parameters $\boldsymbol{\nu}_c$, $\boldsymbol{\eta}$, and $\boldsymbol{\tau}_c$ are estimated in the previous stage, and $\widetilde{\mathbf{S}}$ is a known training matrix, $\mathbf{X}$ is known in Stage~III.} The matrix $\widetilde{\mathbf{D}}(\boldsymbol{\nu}_c)$ shares a similar structural form with ${\mathbf{D}}(\boldsymbol{\nu}_c)$; however, due to the variation in the index $p$ in the third stage, its $p$th row is now defined as $e^{-\mathrm{j} 2\pi (P_s + P_0 + p - 1)\nu}$, evaluated at the corresponding Doppler frequency. Notably, $\widetilde{\mathbf{D}}(\boldsymbol{\nu}_c)$ can be constructed using the Doppler estimates obtained from the previous stage. Subsequently, let $\mathbf{G}_d = \mathrm{diag}(\mathbf{g})$, where the $p$th element of $\mathbf{g}$ is denoted by $g_p$. Accordingly, the $p$th tag message is obtained as
\begin{equation}
    \begin{aligned}
    \hat{g}_p = \frac{\mathbf{Y}_d^*(p,:)\mathbf{X}^T(p,:)}{||\mathbf{X}^T(p,:)||^2}.
    \end{aligned}
    \label{tag-data-estiamtion}
\end{equation}
which can be decoded at the DFRC receiver using the known constellations.

\begin{figure*}[!t]
    \centering
    \begin{subfloat}[]{
        \includegraphics[width=0.48\linewidth]{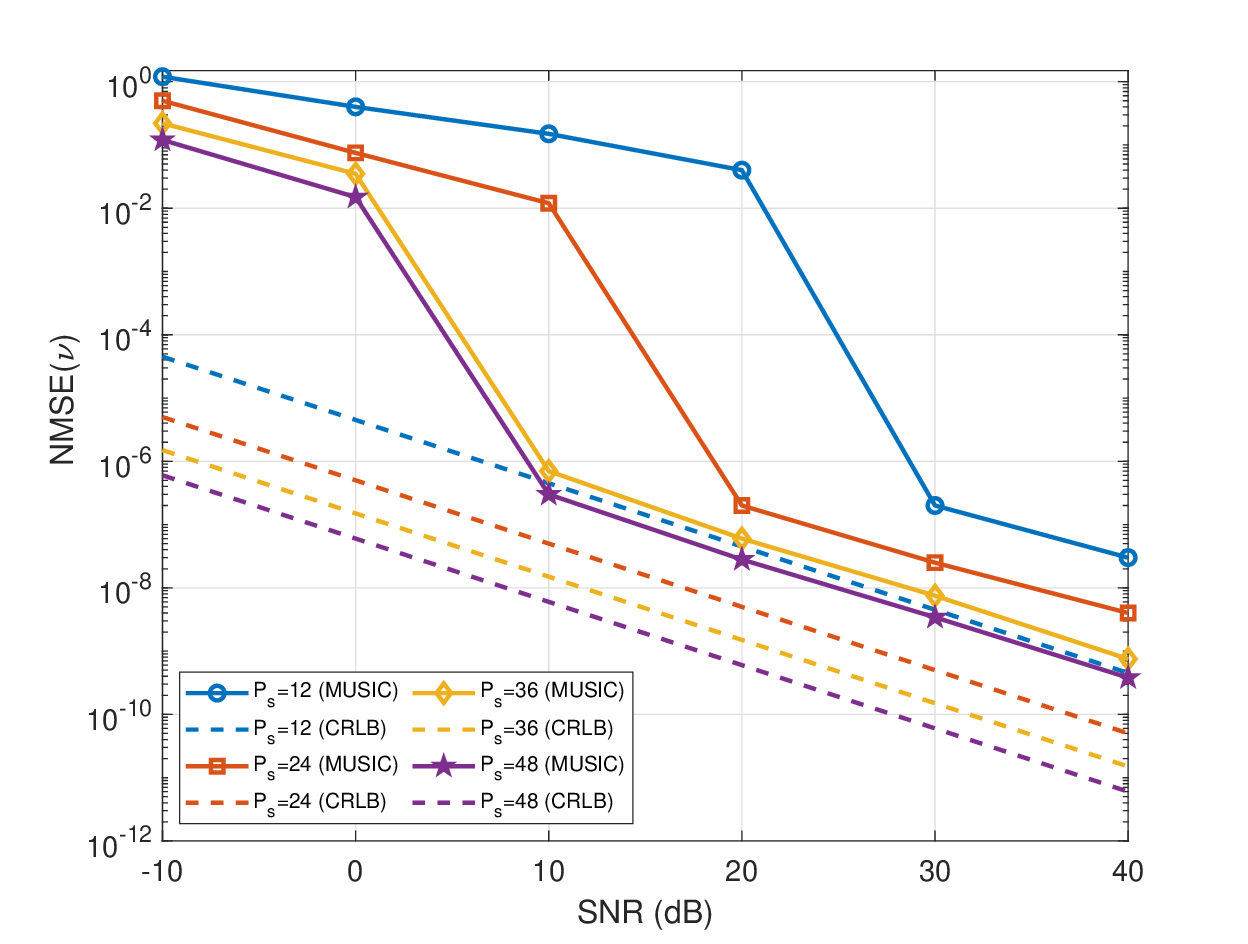}}
        \label{fig:a}
    \end{subfloat}
    \hfill
    \begin{subfloat}[]{
        \includegraphics[width=0.48\linewidth]{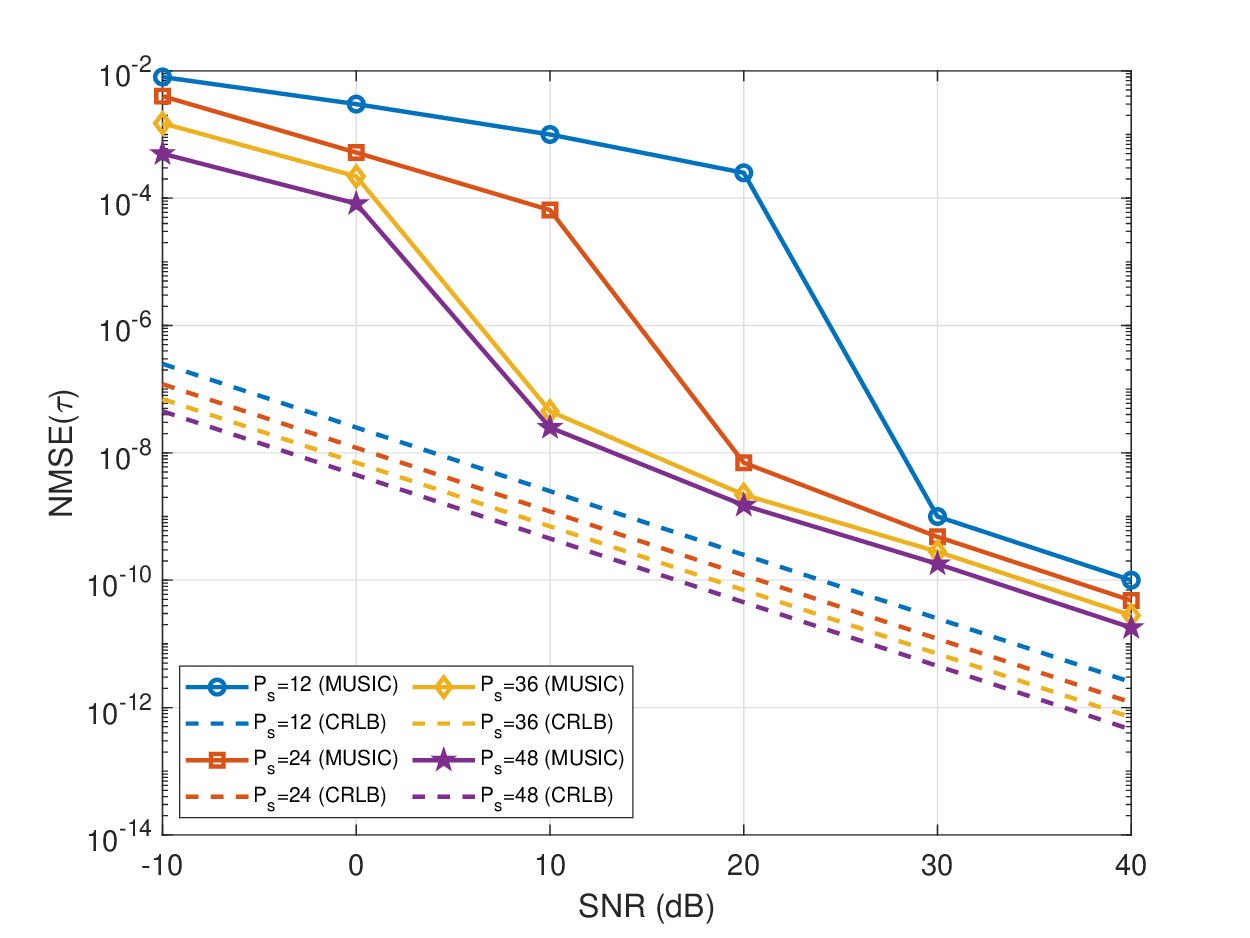}}
        \label{fig:b}
    \end{subfloat}
    
    
    \begin{subfloat}[]{
        \includegraphics[width=0.48\linewidth]{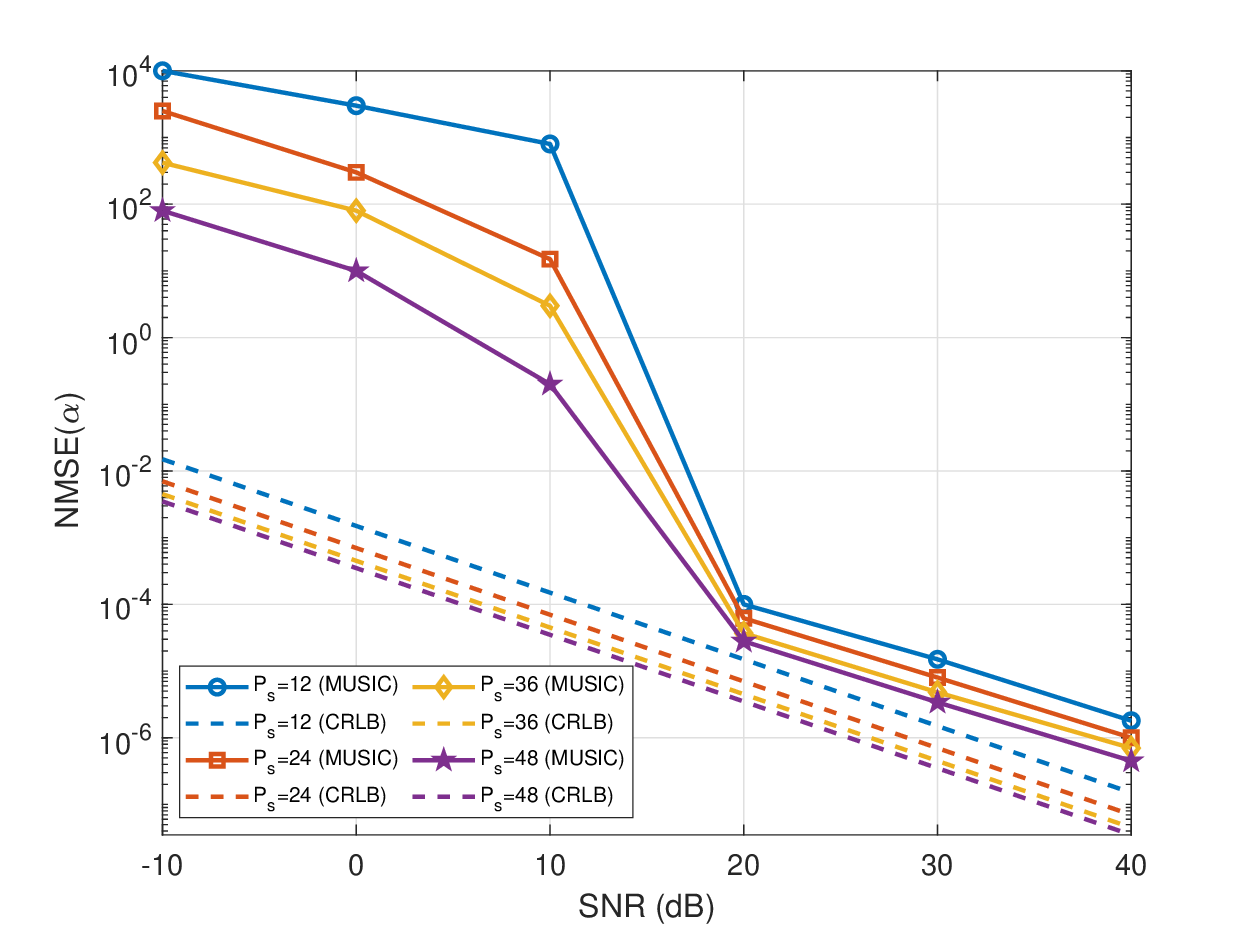}}
        \label{fig:c}
    \end{subfloat}
    \hfill
    \begin{subfloat}[]{
        \includegraphics[width=0.48\linewidth]{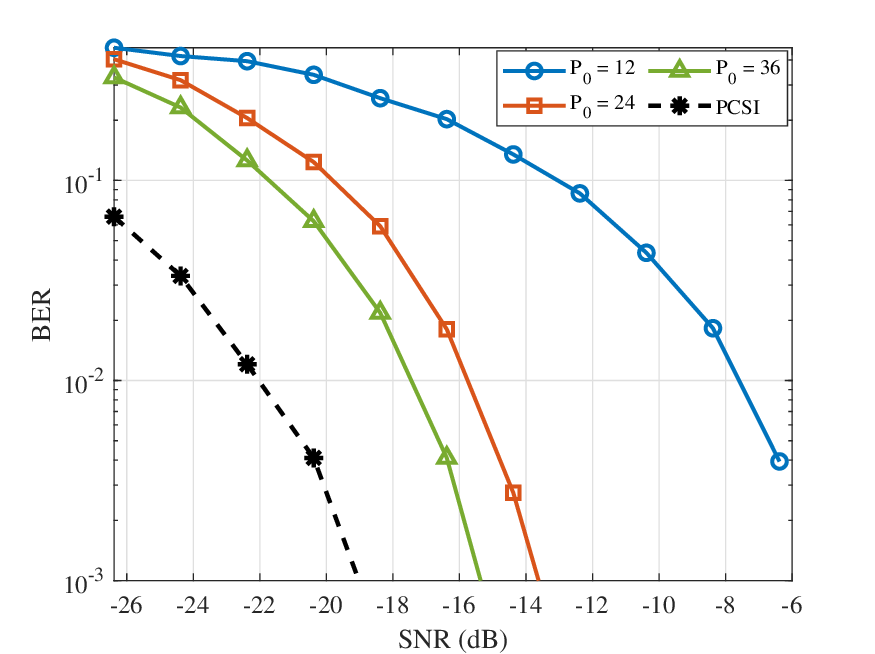}}
        \label{fig:d}
    \end{subfloat}
    
    \caption{(a) NMSE performance of Doppler estimation; 
     (b) NMSE performance of delay estimation; 
             (c) NMSE performance of the reflection coefficient 
             (d) BER performance of tag symbol detection.}
    \label{fig:combined_results}
\end{figure*}

\subsection{Computational complexity}
The computational complexity of performing the SVD on the spatially smoothed observation matrix $\mathbf{Z}$, with $N_s \gg P'N'$, is on the order of $\mathcal{O}\!\left(N_s (P'N')^{2}\right)$. In comparison, extracting the noise subspace from the left singular matrix $\mathbf{U}$ requires only $\mathcal{O}\!\left((P'N')^{2}\right)$ operations. Consequently, the SVD step dominates the overall computational load. In addition, the inversion of the matrix $\boldsymbol{\Psi}$ required for estimating $\boldsymbol{\alpha}_r$ incurs a complexity of $\mathcal{O}(PN\,L^{2})$, where $PN \gg L$. Since Stage~II also employs the 2D-MUSIC algorithm for parameter recovery, its computational burden closely matches that of Stage~I.

\section{Numerical Experiments}
\begin{table}[ht]
\centering
\caption{Simulation Parameters}
\renewcommand{\arraystretch}{1.1}
\setlength{\tabcolsep}{4pt}
\begin{tabular}{lll}
\hline
\textbf{Description} & \textbf{Symbol} & \textbf{Value} \\
\hline
Coherent processing interval & CPI & 10 ms \\
Pulse repetition interval & $T$ & $100\,\mu$s\\
Carrier frequency & $f_c$ & $28$ GHz \\
Total transmitted pulses & $P$ & 100 \\
Number of subcarriers & $N$ & 32 \\
Number of targets & $L$ & 3 \\
Number of clutters & $Q$ & 4 \\
Tag-absorption coefficient & $G$ & 1\\
\hline
\end{tabular}
\label{tab:sim_params}
\end{table}
\textcolor{black}{This section presents simulation results to evaluate the performance of the proposed algorithms for joint estimation of target and clutter parameters, as well as tag symbol detection. The large-scale fading coefficients are modeled as a function of the carrier frequency $f_c$ according to the 3GPP UMi propagation model~\cite{ISBAC_letter}. The noise is modeled as AWGN with variance $\sigma^2 = 10\log_{10}(N_0 B)$ dBm, where $N_0 = -174$ dBm/Hz denotes the thermal noise power spectral density. We consider a coherent processing interval (CPI) of $10$ ms with a pulse repetition interval (PRI) of $100$ $\mu$s, resulting in $100$ pulses per CPI. These pulses are utilized for both target/clutter parameter estimation and tag data detection. The key system and simulation parameters are summarized in Table~I. Unless otherwise specified, the spatial smoothing parameters are set to $P' = P_s/3$ and $N' = 26$.}.

\begin{figure*}[ht]
    \centering
    \setlength{\tabcolsep}{1pt} 
    \renewcommand{\arraystretch}{0} 
   {\includegraphics[width=0.86\linewidth]{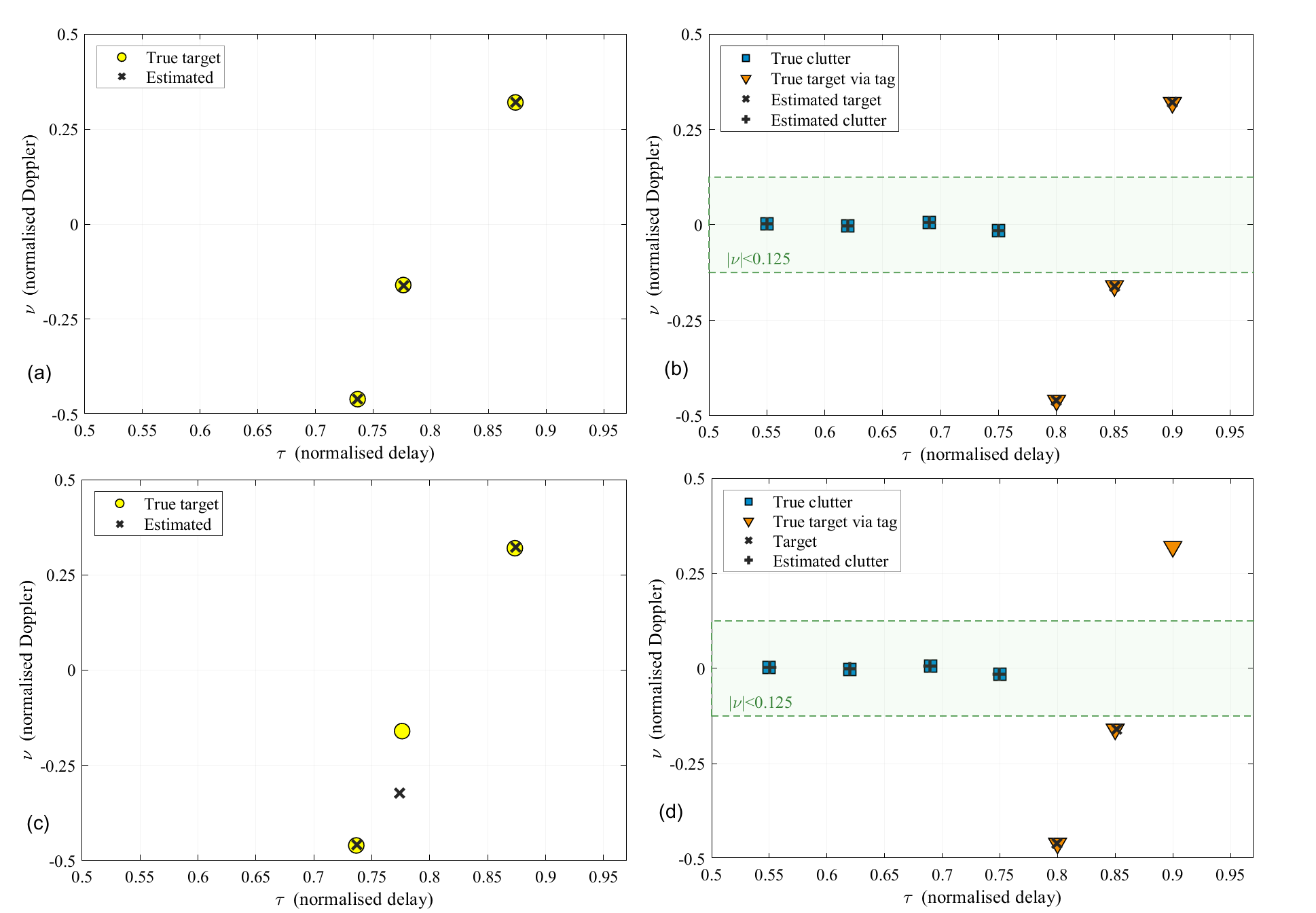}\label{fig:clutter}}\

    \caption{Delay--Doppler estimation performance of the proposed three-stage ISABC framework. (a) Stage-I estimation of three direct-path radar targets at SNR = $10$ dB; (b) Stage-II joint estimation of three tag-assisted target reflections and four clutter components at SNR = $10$ dB; (c) Stage-I estimation of three direct-path radar targets at SNR = $-5$ dB; and (d) Stage-II joint estimation of three tag-assisted target reflections and four clutter components at SNR = $-5$ dB. The shaded region in (b) and (d) ($|\nu|<0.125$) denotes the Doppler threshold used to classify clutter components.}
    \label{fig:grid}
\end{figure*}
\textbf{Target Sensing:} Fig.~3(a)–(c) illustrates the NMSE performance for the Doppler shift, propagation delay, and complex gain estimation, respectively, along with the corresponding CRLBs derived in the Appendix. As expected, increasing the number of sensing pulses $P_s$ enhances the estimation accuracy for all parameters. However, the rate of improvement diminishes as $P_s$ increases. In particular, the NMSE exhibits a substantial reduction as $P_s$ increases from $12$ to $24$, whereas the additional gain from further increasing $P_s$ from $24$ to $36$ becomes marginal, indicating saturation. Furthermore, as shown in Fig.~3(c), the NMSE associated with the complex gain~$\alpha$ shows a higher estimation error than that of the delay and Doppler parameters. This is because $\alpha$ is estimated through an LS procedure after the delay and Doppler parameters are first retrieved using the 2D-MUSIC algorithm. Consequently, any residual estimation error in delay and Doppler propagates into the LS-based gain estimation, leading to a higher NMSE for $\alpha$. 

\textcolor{black}{ In the high-SNR regime, the estimation accuracy is fundamentally limited by the CRLB, which scales inversely with the Fisher information. For the considered spatial-smoothing configuration, the Fisher information grows as $\mathcal{O}(P'^3)$, where $P'=\lfloor P_s/3 \rfloor$. As a result, increasing $P_s$ yields diminishing returns in NMSE reduction, explaining the larger improvement observed when $P_s$ increases from $12$ to $24$ compared to $24$ to $36$. Furthermore, the parallel high-SNR behavior of the NMSE curves with the CRLB indicates that the performance is information-limited rather than constrained by the proposed algorithm.}
 \par
\textcolor{black}{\textbf{Tag Communication:} Fig.~3(d) presents the bit-error-rate (BER) performance for tag data detection and compares it with the idealized case of perfect channel state information (PCSI)The dotted line labeled PCSI corresponds to the performance with Perfect Channel State Information, where the delays, Doppler shifts, and reflection coefficients of all targets and clutters in the Stage-2 system model are assumed to be perfectly known at the receiver. This ideal case serves as a performance benchmark for evaluating the proposed channel estimation approach.}\par 
\textcolor{black}{As expected, the BER decreases with increasing $P_{0}$, since a larger number of Stage-II pulses provides more observations for estimating the composite backscatter channel, thereby improving the accuracy of channel reconstruction and subsequent tag symbol detection.}\par
\textcolor{black}{A behavior similar to the sensing performance is observed. Specifically, increasing $P_{0}$ from $12$ to $24$ yields a significantly larger BER improvement than increasing $P_{0}$ from $24$ to $36$. This trend arises because the estimation accuracy of the Stage-II parameters is fundamentally governed by the available Fisher information, which increases nonlinearly with the effective observation aperture. Consequently, the incremental information gain obtained by increasing $P_{0}$ becomes smaller as $P_{0}$ grows, resulting in diminishing returns in BER performance.}
 \par
\textcolor{black}{\textbf{Delay-Doppler Map:} Fig.~4(a)--(d) presents the estimated delay--Doppler maps obtained in Stages~I and II under two different SNR conditions. Specifically, the top row (Figs.~4(a) and~4(b)) corresponds to a high-SNR scenario ($10$ dB), whereas the bottom row (Figs.~4(c) and~4(d)) corresponds to a low-SNR scenario ($-5$ dB). Figures~4(a) and~4(c) show the Stage-I delay--Doppler estimation of the direct-path target echoes, while Figs.~4(b) and~4(d) depict the Stage-II estimation of the tag-assisted target and clutter components. At $10$ dB, the estimated delay--Doppler pairs closely coincide with the true values, demonstrating accurate recovery of all target and clutter components. At $-5$ dB, although the estimation accuracy degrades due to increased noise, the dominant target and clutter components are still successfully recovered with only a slight increase in estimation error, highlighting the robustness of the proposed framework.}

\textcolor{black}{It should be noted that the delay and Doppler parameters of the tag-assisted target echoes differ from those estimated in Stage~I because the additional propagation through the tag introduces a different propagation delay, Doppler shift, and complex channel gain. Consequently, the Stage-II target components cannot be directly associated with their Stage-I counterparts based solely on the estimated delay--Doppler parameters.}

\textcolor{black}{Nevertheless, target and clutter components can be reliably discriminated by exploiting their inherent physical Doppler characteristics. In UMi, clutters, such as buildings and infrastructure, produces nearly zero Doppler, while slowly moving clutter (e.g., vegetation swayed by wind) exhibits only small Doppler frequencies. In contrast, moving targets generate significantly larger Doppler shifts owing to their higher radial velocities. Accordingly, the shaded region in Figs.~4(b) and~4(d), corresponding to $|\nu|<0.125$, is designated as the clutter region. Components whose estimated Doppler lies within this region are classified as clutter, whereas those outside the region are identified as target-assisted reflections. As evident from the figure, this simple physics-based criterion enables reliable target--clutter discrimination under both high- and low-SNR conditions.}

\textcolor{black}{\textbf{Comparison with the SotA schemes:} 
Fig.~5 compares the sensing NMSE of the proposed off-grid 2D-MUSIC algorithm with the 2D periodogram, 2D-OMP, and conventional on-grid 2D-MUSIC for delay and Doppler estimation. It is observed that both the periodogram and 2D-OMP exhibit early error saturation due to limited resolution and grid mismatch, while the on-grid 2D-MUSIC achieves improved accuracy but remains constrained by grid quantization at high SNR. In contrast, the proposed off-grid 2D-MUSIC, aided by golden-section refinement, effectively mitigates the grid quantization error and continues to improve with increasing SNR, closely following the CRLB trend. At high SNR, the proposed method achieves approximately four orders of magnitude lower NMSE than the best grid-based approach, demonstrating the effectiveness of off-grid refinement with only a marginal increase in computational complexity.}

 \textcolor{black}{\textbf{Trade-off:} Fig.~5 illustrates the sensing-communication trade-off in the considered ISABC system at SNR levels of $10$~dB and $20$~dB. The total pulse budget $P$ and the tag data-bearing pulses $P_c$ are kept fixed to enforce a constant tag data rate, while the sensing allocation $P_s$ is varied. The spatial smoothing aperture is adaptively set as $P' = P_s/5$. As $P_s$ grows, the NMSE of the target complex gain~$\alpha$ decreases (i.e., estimation accuracy improves) owing to the increased number of independent sensing snapshots. In contrast, the tag BER exhibits controlled degradation, stemming from the reduced availability of $P_0$ pulses dedicated to estimating and reconstructing the equivalent bistatic channel used for tag demodulation in Stage~III. Since channel estimation accuracy scales with $P_0$, its reduction amplifies residual channel uncertainty, which propagates to the coherent detection of tag-modulated symbols, leading to higher decoding error probability.}

\begin{figure*}[t]
\centering

\begin{minipage}[t]{0.49\textwidth}
    \centering
    \includegraphics[width=\linewidth]{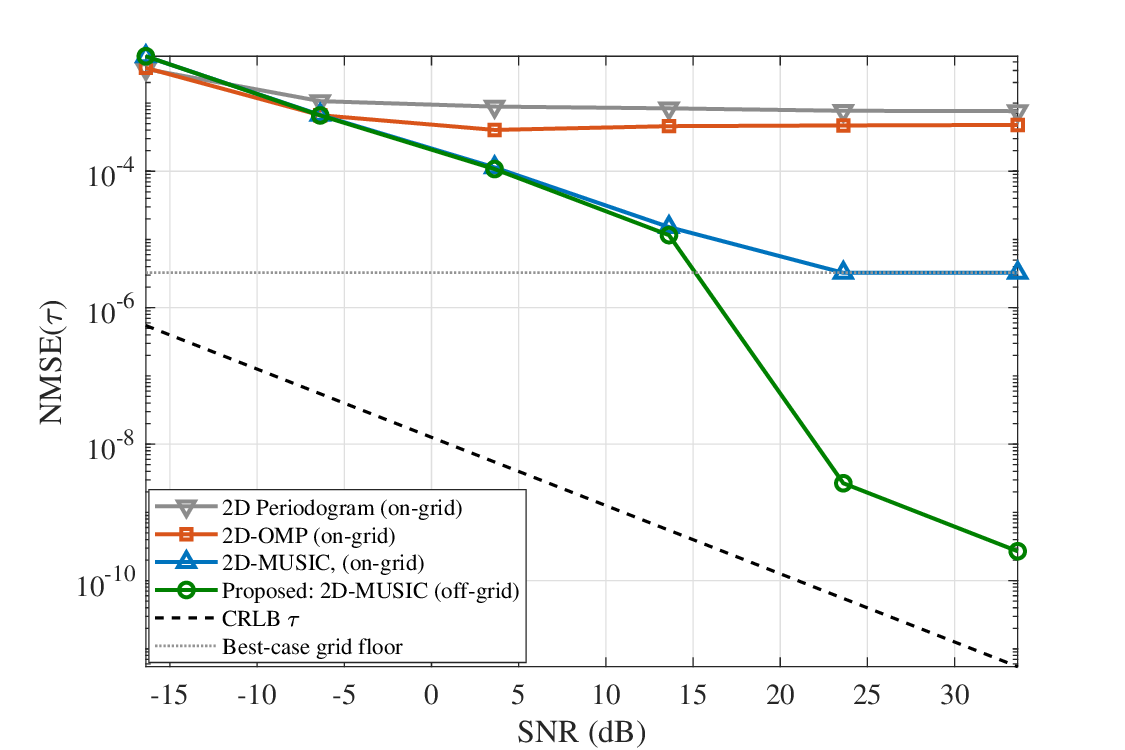}
\end{minipage}\hfill
\begin{minipage}[t]{0.49\textwidth}
    \centering
    \includegraphics[width=\linewidth]{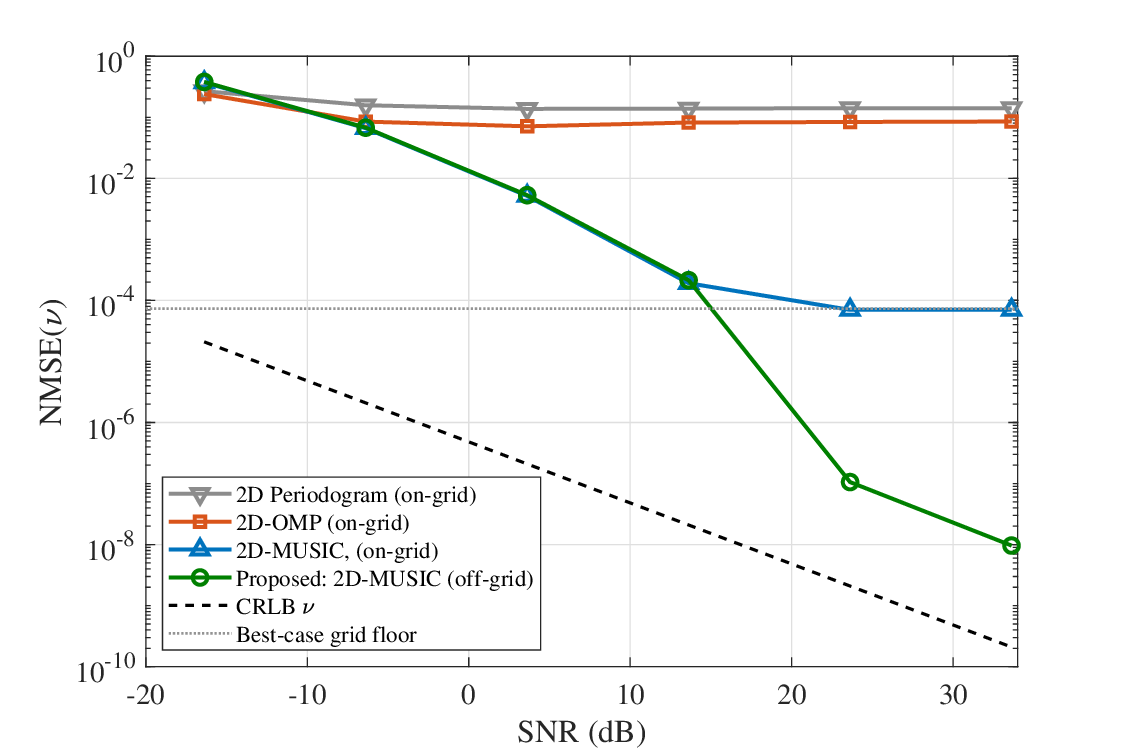}
\end{minipage}

\caption{Sensing NMSE comparison with state-of-the-art algorithms: (a) delay estimation; (b) Doppler estimation.}
\label{fig:sota}
\end{figure*}

\section{Summary}
This work has developed a unified, practical framework for joint target and clutter parameter estimation, along with tag symbol detection, for ISABC systems at a bistatic DFRC receiver. A complete end-to-end signal model and a new pulse frame structure were proposed to support simultaneous sensing and backscatter communication. The three-stage estimation and detection strategy has shown strong capability in reliably extracting target and clutter parameters, including the complex gain, delay, and Doppler, and further enables dependable tag data decoding using the reconstructed equivalent channel. The performance of the proposed solution was rigorously verified through simulations, and benchmarked against CRLB. In addition, the sensing–communication interaction was quantitatively characterized, showing that dedicating more pulses to sensing improves parameter estimation accuracy but limits the pulses used for channel probing, which is essential for backscatter demodulation—leading to increased tag detection errors. These observations provide clear system design insights and guidelines for pulse allocation, offering a balanced performance trade space for future ISABC receiver architectures.

\appendix
\section{CRLB Derivation for Sensing Parameters}
Consider the received signal model in \eqref{eq:2D_MUSIC_system_model2}. Vectorizing the received signal column-wise yields:
\begin{equation}
\mathbf{y} = \vecop(\mathbf{Y})
           = \mathbf{\Phi}(\boldsymbol{\tau},\boldsymbol{\nu})\,\boldsymbol{\alpha} 
           + \mathbf{e}
\end{equation}
where $\mathbf{y}\in\mathbb{C}^{PN\times 1}$, $\mathbf{e}\sim \mathcal{CN}(\mathbf{0},\sigma^2\mathbf{I})$, and the  $\mathbf{\Phi}\in\mathbb{C}^{PN\times L}$ has elements
\begin{equation}
\Phi_{(n-1)P+p,l}
= s_n e^{j2\pi\nu_l t_p} e^{-j2\pi f_n \tau_l}.
\label{eq:PhiElem}
\end{equation}

Let $\boldsymbol{\phi}_l$ denote the $l$-th column of $\mathbf{\Phi}$.  
Define the parameter vector
\begin{equation}
\boldsymbol{\theta}
=
[\tau_1,\dots,\tau_L,\;
 \nu_1,\dots,\nu_L,\;
 \alpha_{R,1},\dots,\alpha_{R,L},\;
 \alpha_{I,1},\dots,\alpha_{I,L}]^{T}
\end{equation}
where $\alpha_l=\alpha_{R,l}+j\alpha_{I,l}$. Define the diagonal matrices
\begin{align}
\mathbf{F} &= \diagop(
f_1 \mathbf{1}_P,\;
f_2 \mathbf{1}_P,\;\ldots,\;
f_N \mathbf{1}_P), \\
\mathbf{T} &= \diagop(
1,\ldots,P_s,\;
1,\ldots,P_s,\;\ldots,\;
1,\ldots,P_s),
\end{align}
where each $f_n$ is repeated $P_s$ times in $\mathbf{F}$, and each pulse index $p=1,\hdots,P_s$ is repeated $N$ times in $\mathbf{T}$. The noiseless mean is $\mathbf{h}=\mathbf{\Phi}\boldsymbol{\alpha}$.  
Using \eqref{eq:PhiElem}, the derivatives are
\begin{equation}
\begin{aligned}
\frac{\partial\mathbf{h}}{\partial\tau_k} &= -j2\pi \alpha_k \mathbf{F}\boldsymbol{\phi}_k, &
\frac{\partial\mathbf{h}}{\partial\nu_k} &= j2\pi \alpha_k \mathbf{T}\boldsymbol{\phi}_k, \\
\frac{\partial\mathbf{h}}{\partial \alpha_{R,k}} &= \boldsymbol{\phi}_k, &
\frac{\partial\mathbf{h}}{\partial \alpha_{I,k}} &= j\boldsymbol{\phi}_k.
\end{aligned}
\end{equation}
Collecting all derivatives, define the Jacobian
\begin{equation}
\mathbf{G}
=
\big[\,\mathbf{G}_\tau\;\; \mathbf{G}_\nu\;\;
       \mathbf{G}_{\alpha_R}\;\;
       \mathbf{G}_{\alpha_I}\,\big]
\end{equation}
with
\begin{equation*}
\begin{aligned}
\mathbf{G}_\tau &= -j2\pi\,\mathbf{F}\,\mathbf{\Phi}\,\diagop(\boldsymbol{\alpha}), & 
\mathbf{G}_\nu  &=  j2\pi\,\mathbf{T}\,\mathbf{\Phi}\,\diagop(\boldsymbol{\alpha}), \\
\mathbf{G}_{\alpha_R} &= \mathbf{\Phi}, & 
\mathbf{G}_{\alpha_I} &= j\,\mathbf{\Phi}.
\end{aligned}
\end{equation*}
For real parameters under circular complex Gaussian noise, the Fisher information matrix (FIM) is defined as
\begin{equation}
\mathbf{I}(\boldsymbol{\theta})
=
\frac{2}{\sigma^{2}}
\Re\!\left(\mathbf{G}^{H}\mathbf{G}\right).
\label{eq:FIMbase}
\end{equation}
Define the Gram-type matrices
\begin{alignat*}{3}
\mathbf{\Lambda} &= \mathbf{\Phi}^{H}\mathbf{\Phi}, &\quad
\mathbf{\Lambda}_{F} &= \mathbf{\Phi}^{H}\mathbf{F}\mathbf{\Phi}, &\quad
\mathbf{\Lambda}_{T} &= \mathbf{\Phi}^{H}\mathbf{T}\mathbf{\Phi}, \\
\mathbf{\Lambda}_{F^2} &= \mathbf{\Phi}^{H}\mathbf{F}^{2}\mathbf{\Phi}, &\quad
\mathbf{\Lambda}_{T^2} &= \mathbf{\Phi}^{H}\mathbf{T}^{2}\mathbf{\Phi}, &\quad
\mathbf{\Lambda}_{TF} &= \mathbf{\Phi}^{H}\mathbf{T}\mathbf{F}\mathbf{\Phi}.
\end{alignat*}
Let $\mathbf{A}=\diagop(\boldsymbol{\alpha})$.  
Substituting into \eqref{eq:FIMbase} yields
\begin{equation}
\resizebox{\columnwidth}{!}{$
\mathbf{I}(\boldsymbol{\theta}) =
\frac{2}{\sigma^{2}}
\begin{bmatrix}
4\pi^{2}\mathbf{A}^{H}\mathbf{\Lambda}_{F^{2}}\mathbf{A}
& -4\pi^{2}\mathbf{A}^{H}\mathbf{\Lambda}_{TF}\mathbf{A}
& -2\pi\,\mathrm{Im}\!\left(\mathbf{A}^{H}\mathbf{\Lambda}_{F}\right)
& -2\pi\,\mathrm{Re}\!\left(\mathbf{A}^{H}\mathbf{\Lambda}_{F}\right)
\\[1ex]
-4\pi^{2}\mathbf{A}^{H}\mathbf{\Lambda}_{TF}\mathbf{A}
& 4\pi^{2}\mathbf{A}^{H}\mathbf{\Lambda}_{T^{2}}\mathbf{A}
& 2\pi\,\mathrm{Im}\!\left(\mathbf{A}^{H}\mathbf{\Lambda}_{T}\right)
& 2\pi\,\mathrm{Re}\!\left(\mathbf{A}^{H}\mathbf{\Lambda}_{T}\right)
\\[1ex]
-2\pi\,\mathrm{Im}\!\left(\mathbf{A}^{H}\mathbf{\Lambda}_{F}\right)^{\top}
& 2\pi\,\mathrm{Im}\!\left(\mathbf{A}^{H}\mathbf{\Lambda}_{T}\right)^{\top}
& \mathrm{Re}(\mathbf{\Lambda}) & \mathbf{0}
\\[1ex]
-2\pi\,\mathrm{Re}\!\left(\mathbf{A}^{H}\mathbf{\Lambda}_{F}\right)^{\top}
& 2\pi\,\mathrm{Re}\!\left(\mathbf{A}^{H}\mathbf{\Lambda}_{T}\right)^{\top}
& \mathbf{0} & \mathrm{Re}(\mathbf{\Lambda})
\end{bmatrix}
$}
\end{equation}
The CRLB follows from the FIM inverse \cite{SMKay},
\begin{equation}
\mathrm{CRLB}(\boldsymbol{\theta})
=
\mathbf{I}^{-1}(\boldsymbol{\theta}).
\end{equation}

 \begin{figure}[t]
    \centering
    \includegraphics[width=0.8\linewidth]{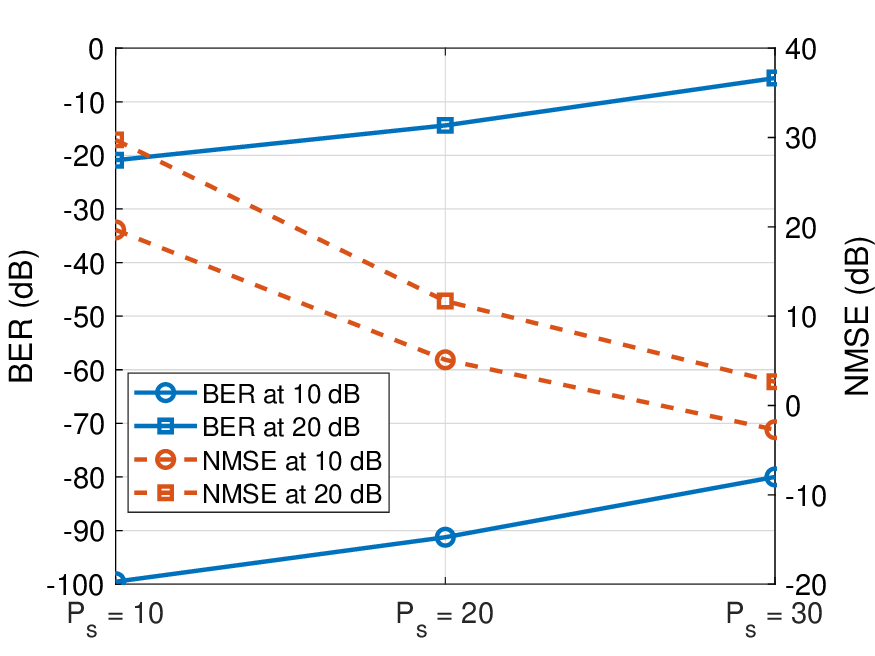}
    \caption{ Sensing and communication trade-off while increasing $P_s$ with fixed $P=100$ and $P_c=50$.}
    \label{fig:placeholder}
\end{figure}

\bibliographystyle{IEEEtran}
\bibliography{biblo}

\end{document}